\documentclass[lineno]{article}

\usepackage[utf8]{inputenc}
\usepackage{graphicx,lineno}
\usepackage{amsfonts}
\usepackage{longtable}
\usepackage{subcaption}
\usepackage{xcolor,colortbl}
\usepackage{makecell}
\usepackage{url,lineno}
\usepackage{tikz}
\usepackage{diagbox}
\usepackage{float}
\usepackage{makecell}
\usepackage[normalem]{ulem}
\usepackage[labelfont=bf, font=scriptsize]{caption}
\usepackage{placeins}
\usepackage{authblk}
\usepackage{array}
\usepackage{multirow}
\usepackage[a4paper, total={6.3in, 10in}]{geometry}
\usepackage[colorlinks=true, citecolor=blue, filecolor=blue, linkcolor=black]{hyperref}
\usepackage{bbm}
\usepackage{color}
\usepackage{epstopdf}
\usepackage{stmaryrd}
\usepackage{comment}
\usepackage{multicol}
\usepackage{amsmath,amssymb} 
\usepackage{color}
\usepackage{mathtools}
\usepackage{lineno}
\usepackage[T1]{fontenc}    
\usepackage{url}            
\usepackage{booktabs}       
\usepackage{amsfonts}       
\usepackage{nicefrac}       
\usepackage{microtype}      
\usepackage{lipsum}
\usepackage{xcolor,colortbl}
\usepackage{amsmath}
\usepackage[square,numbers]{natbib}

\usepackage{lastpage,fancyhdr}

\newcommand{\be}{\begin{equation}}
\newcommand{\ee}{\end{equation}}
\newcommand{\bea}{\begin{eqnarray}}
\newcommand{\eea}{\end{eqnarray}}

\definecolor{LightCyan}{rgb}{1,1,1}
\definecolor{lightblue}{rgb}{1,1,1}
\definecolor{lesslightblue}{rgb}{1,1,1}
\definecolor{bb}{rgb}{1,1,1}
\definecolor{bbb}{rgb}{1,1,1}

\definecolor{rr}{rgb}{1,1,1}
\definecolor{rrr}{rgb}{1,1,1}

\definecolor{Gray}{rgb}{1,1,1}
\definecolor{DGray}{rgb}{1,1,1}
\definecolor{VDGray}{rgb}{1,1,1}

\newcolumntype{g}{>{\columncolor{Gray}}l}
\newcolumntype{j}{>{\columncolor{DGray}}l}
\newcolumntype{h}{>{\columncolor{white}}p{1cm}}
\newcolumntype{o}{>{\columncolor{bbb}}p{1cm}}

\newcolumntype{y}{>{\columncolor{rrr}}p{1cm}}
\newcolumntype{u}{>{\columncolor{rr}}p{1cm}}

\newcolumntype{v}{>{\columncolor{bb}}c}
\newcolumntype{m}{>{\columncolor{bb}}p{1.6cm}}
\newcolumntype{b}{>{\columncolor{bbb}}p{1.6cm}}

\newcolumntype{r}{>{\columncolor{rrr}}p{1.5cm}}
\newcolumntype{t}{>{\columncolor{rr}}p{1.5cm}}

\newcolumntype{q}{>{\columncolor{Gray}}p{1.0cm}}
\newcolumntype{w}{>{\columncolor{DGray}}p{1.0cm}}

\newcolumntype{Q}[1]{>{\columncolor{Gray}}p{#1}}
\newcolumntype{W}[1]{>{\columncolor{DGray}}p{#1}}

\newcolumntype{z}{>{\columncolor{Gray}}p{10cm}}

\definecolor{lime}{HTML}{A6CE39}
\DeclareRobustCommand{\orcidicon}{%
	\begin{tikzpicture}
	\draw[lime, fill=lime] (0,0) 
	circle [radius=0.16] 
	node[white] {{\fontfamily{qag}\selectfont \tiny ID}};
	\draw[white, fill=white] (-0.0625,0.095) 
	circle [radius=0.007];
	\end{tikzpicture}
	\hspace{-2mm}
}

\foreach \x in {A, ..., Z}{%
	\expandafter\xdef\csname orcid\x\endcsname{\noexpand\href{https://orcid.org/\csname orcidauthor\x\endcsname}{\noexpand\orcidicon}}
}

\begin{document}  

\title{\textbf{Risk mapping novel respiratory pathogens with large-scale dynamic contact networks}}

\author[1]{Matthijs Romeijnders\orcidA{}}
\affil[1]{Department of Information and Computing Sciences, Utrecht University, Utrecht, The~Netherlands}
\author[2]{Michiel van Boven\orcidB{}\footnote{Corresponding author: r.m.vanboven-2@umcutrecht.nl}}
\affil[2]{Julius Center for Health Sciences and Primary Care, Utrecht University, Utrecht, The~Netherlands} 
\author[1]{Debabrata~Panja\orcidC{}}

\maketitle 

\section*{Abstract}
 
{\bf Background:} Human-to-human transmission of pathogens fundamentally depends on interactions among infectious and susceptible individuals, yet traditional population-scale models often overlook the stochastic, behaviour-driven, and highly heterogeneous nature of these interactions. 

\noindent {\bf Methods:} Here, we develop a large-scale actor-based model capturing early epidemic dynamics of a novel respiratory pathogen on dynamic contact networks. We build these networks upon explicitly integrating detailed demographic and residential registry data from the Netherlands. The model simulates the Dutch population characterised by age, residency and mobility patterns, with actors interacting stochastically across households, workplaces and schools. 

\noindent {\bf Results:} We show how the geographic and demographic profiles of initial cases impact transmission trajectories, with densely populated municipalities in the country’s western core acting as key hubs driving epidemic spread. The framework enables rigorous assessment of intervention strategies incorporating behavioural adaptations. As case studies, we quantify the effects of symptomatic self-isolation and travel restrictions to and from major urban centres, highlighting their potential to modulate epidemic outcomes. 

\noindent {\bf Conclusions:} Our findings underscore the necessity of integrating fine-scale human-to-human contact realism and population scale in epidemic forecasting and control.

\section*{Plain-language summary}

Mathematical modelling of infectious diseases is a cornerstone for understanding and predicting how pathogens spread in populations. Current models of disease spread, despite their widespread use, rely on  one-size-fits-all assumptions that fail to capture the dynamic, and adaptive nature of real-world human interactions. Network models have the fine detail needed to represent these complexities, but face challenges in scalability and generalisability. Here, we introduce a novel hybrid model that  combines the realism of network models with the adaptability of population-level models, enabling a more accurate overall analysis. Our framework advances epidemic modelling by bridging detailed interpersonal behaviour and large-scale generalisability.

\section*{Introduction}

Respiratory pathogens such as influenza A, respiratory syncytial virus (RSV), {\it Bordetella pertussis}, or {\it Streptococcus pneumoniae} are major health threats. The recent addition of SARS-CoV-2 to the list, in terms of its sudden emergence, rapid progression through the world, the trail of suffering and death, and the strain on healthcare systems, has introduced a new layer of urgency, underscoring the need for being prepared for pandemics arising from novel respiratory pathogens.

\vspace{2mm}
\noindent Epidemiological modelling can inform both the risks and the effectiveness of interventions, but a key challenge that modelling faces is that the spread of any human-to-human pathogen depends on the underlying contact network, which is shaped by many factors, for instance how people travel, mix, and interact. Even within a single region, the contact network will be stochastic, dynamic, and heterogeneous across space and time. Micro-level contact heterogeneities drive macro-scale variability in infection and transmission risks, which has important implications for designing targeted interventions and understanding how these measures may differentially affect various population groups, for instance, due to varying contact intensities \cite{mossongetal2008}. Despite this understanding, the study of infectious disease transmission has traditionally relied on compartmental models with rate equations described by ODEs (henceforth compartmental models), which have provided important insights into the dynamics of infectious diseases but assume homogeneous mixing within the compartments. Populations in these models are typically subdivided into discrete compartments such as susceptible, exposed (i.e. infected but not yet infectious), infectious, and recovered. In this setting, the epidemic dynamics can be described by differential equations, with individuals moving between the compartments based on the prescribed rates of transmission, recovery, and other processes \cite{diekmann2013}. Common extensions to these models include some heterogeneities, such as age structure and sex specificity \cite{diekmann2013, rozhnova2020}, but they still adhere to the principle of homogeneous mixing within the compartments.

\vspace{2mm}
\noindent To overcome some of these limitations, metapopulation or patch-based models incorporating the spatial structure of populations have been developed in epidemiology, adopted mainly from ecology \cite{grenfell1995, hanski1999}. In metapopulation models, individuals are grouped into subpopulations, each of them again modelled as a well-mixed compartment. Interactions between subpopulations are then added through migration or travel through rate equations, capturing some of the spatial heterogeneity \cite{riley2007, colizza2007, colizza2008, balcan2009, belik2011, apolloni2014, gomezgardenes2018, hazarie2021, cota2021, zunker2024,gosgens2021, uiterkamp2022}. Despite this added layer of sophistication, metapopulation models still often rely on fixed mixing rates within large subpopulations (see \cite{colizza2007} for an exception), leaving out the stochastic, dynamic, and heterogeneous nature of real-world human-to-human contacts. Recent examples include applications to the early dynamics of SARS-CoV-2 in the Netherlands \cite{gosgens2021, uiterkamp2022}.

\vspace{2mm}
\noindent Prompted by developments in network science, explicit representation of individual humans as nodes and contacts amongst them as links has emerged as a highly granular alternative for epidemiological modelling. Starting originally with static network models \cite{Keeling2005}, the field of network epidemiology has developed fast over the past decades with the progressive incorporation of the dynamic aspects of human-to-human contact networks \cite{masuda2017}. Overarchingly, network models are able to provide a more detailed and realistic picture of pathogen transmission than compartmental models, by incorporating the heterogeneity of contact patterns, capturing the effects of super-spreaders, clustering, and network structure on outbreak dynamics. In network epidemiology, static network models have been used extensively to study the dynamics of infectious diseases, for instance, to quantify transmission of (pandemic) pathogens at a global scale, often using the global aviation network as a proxy for inter-country mobility \cite{Keeling2005, hufnagel2004, brockmann2006, bajardi2011, grady2012, brockmann2013}. However, static network models assume that contact patterns remain fixed over time, which is unrealistic, since human interaction patterns are not only dynamic but are also adaptive due to behavioural changes and public health interventions. In light of this, dynamic network models are becoming an increasingly popular choice for epidemiological modelling, yet they have been used in limited scopes, e.g., to analyse specific past instances of pathogen spreading \cite{masuda2017, brockmann2013, liu2022, belik2011}. The lack of relevant real-world granular-level data and significant computation times for scaling up model projections to population-level are significant bottlenecks for dynamic network epidemiology to attain the status of a competitive, predictive, generalisable and realistic alternative for modelling epidemiological dynamics.

\vspace{2mm}
\noindent In parallel, another line of work has focused on the development of large-scale agent-based models (ABMs). These have also improved upon compartmental and static-network models by representing individuals and heterogeneous contact settings. For instance, FluTE and EpiCast simulate US-scale influenza dynamics at daily time-steps with aggregated contacts in synthetic communities, but do not incorporate adaptive behaviour beyond stylised interventions \cite{germann2006, chao2010}. Others, such as FRED, Covasim, OpenABM-Covid19, JUNE and others implement households, schools, workplaces, and communities across larger regions, but likewise rely on daily time steps and highly abstracted behavioural adaptations \cite{grefenstette2013, kerr2021, hinch2021, aylettbullock2021, hoertel2020}. Taken together, existing ABMs have advanced our understanding of epidemic spread and interventions, but none combine population-scale coverage, fine spatiotemporal granularity, and realistic behavioural adaptation.

\vspace{2mm}
\noindent Here, using the Netherlands as a case study, we present a dynamic network model to provide a comprehensive analysis of spreading risks, transmission patterns over time and in space, and the effectiveness of interventions when a respiratory pathogen is introduced in any Dutch municipality. Importantly, by virtue of taking a hybrid approach, the model allows us to scale up the epidemiological analysis to the population level while also retaining important fine-scale spatiotemporal details. The spatial structure at municipality-level resolution includes the demographic stratification of the population represented by actors at 1:100 population ratio, and resembles that of a metapopulation model. Next to that, embracing the spirit of network epidemiology, human-to-human contacts are built stochastically and dynamically within the embedding of spatial and demographic stratification, drawn at an hourly resolution from probability distributions that are calibrated to across-municipality mobility and local-settings mixing patterns such as home, school and work (Fig. \ref{fig:model_schematics} and model flowchart in Results section, further elaborated in Methods). Transmission of respiratory pathogens takes place stochastically over this dynamic contact network, obeying standard infection dynamics of latency, infectiousness and recovery (see Methods for model and parameterisation). This hybrid approach not only allows us to capture the spatial structure of the population but it also allows for incorporating the temporal variability in contact patterns, leading to more realistic predictions of disease spread and effectiveness of interventions. Moreover, conformation to large-scale mobility and mixing patterns ensures generalisability of our findings within the (cultural) context of the population. An earlier edition of this model was used --- including calibration and validation --- to analyse the first COVID-19 pandemic wave dynamics in the Netherlands \cite{dekker2023}. 

\section*{Methods}

The following steps summarise the core mechanisms underlying our model. These steps yield a full dynamic network of actor contacts, denoted as events in Fig. \ref{fig:model_schematics}. As pointed out in the introduction, An earlier edition of this model was used --- including calibration and validation --- to analyse the first COVID-19 pandemic wave dynamics in the Netherlands \cite{dekker2023}. These calibration steps were repeated with the current model in Ref. \cite{larsthesis}. 

\subsection*{Step 1: Demographic and residential stratification of the actors}

One of the aims of the model was to accurately reflect the demographic composition of the Dutch population. To achieve this, the model employed a 1:100 scale, representing the 2019 population of 17.34 million with approximately 170,000 actors. Since each municipality had to contain whole-numbered actors in every demographic group, the total number of actors fell slightly below 173,400. Using age and occupation data obtained from the Dutch Central Bureau of Statistics (CBS), actors were classified into one of eleven demographic groups so that the resident actors of each municipality reflected the observed age and occupation distribution. A breakdown of the demographic groups is shown in Tab.~\ref{tab:demographic_groups}.

\begin{table}[hpt!]
\centering
\renewcommand{\cellalign}{cl}
\renewcommand{\arraystretch}{1.5}
\setlength{\tabcolsep}{2pt}
\begin{tabular}{|p{2.5cm}||p{1.3cm}|q|p{1.2cm}||p{1.8cm}|p{2.4cm}|m|p{1.8cm}||}

\hline
\multicolumn{1}{|l||}{\textbf{Group}} & \multicolumn{3}{j||}{\textbf{Criteria}} & \multicolumn{4}{v||}{\textbf{Attributes}} \\
\hline
\hline
\textbf{} & \textbf{Age (y)} & \textbf{Work} & \textbf{School} & \textbf{\makecell{National\\ total\\ (fraction)}} & \textbf{\makecell{Av. time not\\ in home\\ municipality}} & \textbf{\makecell{Daytime\\ mixing}} & \textbf{\makecell{Nighttime\\ mixing}}\\

\hline
\makecell{Pre-school\\ children} & 0–4 & - & - & \makecell{851,880\\ (4.9\%)} & 6\% & Home   & Home  \\
\hline
\makecell{Primary school\\ children} & 5–11 & - & Yes & \makecell{1,295,380\\ (7.5\%)} & 6\% & School & Home \\
\hline
\makecell{Secondary\\ school children} & 12–16 & - & Yes & \makecell{991,290\\ (5.7\%)} & 6\% & School & Home \\
\hline
Students & 17–24 & - & Yes & \makecell{1,086,240\\ (6.2\%)} & 26\% & \makecell{School \&\\ Work} & Home \\
\hline
\makecell{Non-studying\\ adolescents} & 17–24 & - & - & \makecell{632,530\\ (3.6\%)} & 26\% & Work & Home \\
\hline
\makecell{Middle-age\\ working} & 25–54 & Yes & - & \makecell{5,530,360\\ (31.8\%)} & 26\% & Work & Home \\
\hline
\makecell{Middle-age\\ unemployed} & 25–54 & - & - & \makecell{1,231,780\\ (7.1\%)} & 6\% & Home   & Home  \\
\hline
\makecell{Higher-age\\ working} & 55–67 & Yes & - & \makecell{1,623,040\\ (9.3\%)} & 26\% & Work & Home  \\
\hline
\makecell{Higher-age\\ unemployed} & 55–67 & - & - & \makecell{1,109,170\\ (6.4\%)} & 6\% & Home   & Home \\
\hline
Elderly & 68–80 & - & - & \makecell{2,102,530\\ (12.2\%)} & 6\% & Home   & Home  \\
\hline
Eldest & 80+ & - & - & \makecell{802,670\\ (4.6\%)} & 6\% & Home   & Home  \\
\hline

\end{tabular}
\caption{Demographic groups and associated attributes. Daytime and nighttime mixing refers to which mixing matrix is used during day- or nighttime. } 
\label{tab:demographic_groups}
\end{table}

\subsection*{Steps 2 and 3: Actor Mobility}

The model was designed to capture the complexity of real-life travel patterns. To do so, a weekly travel schedule was generated for each actor, with distinct daily patterns that repeated every week. The hourly movements of actors were sampled from Dirichlet distributions which used normalised gravity model movements as scale parameters. Daily travel patterns were constructed under the assumption that actors are primarily away from home during mornings and afternoons. For example, if an actor spends 15 hours at home, they are assumed to be at home from 00:00–7:30 and 17:30–24:00, splitting the total equally between the start and end of the day. Time spent in other municipalities was then placed in the leftover time in a random order. 

\vspace{2mm}
\noindent For each actor a travel schedule was generated depending on its demographic characteristics. For example, it was assumed that working adults and students spend 25$\%$ (42 hours per week) of their time in other municipalities, whereas other groups only spend 5$\%$ (8.4 hours per week) of their time away from home. This assumption was incorporated by appropriate weighting of the Dirichlet distributions. In addition to this, actors' movements were sampled by using one of two sets of weights. Firstly, for frequent trips to work or school a typical gravity model was used with the gravitational constant $G = 0.5$ to account for double counting for return journeys. The weight for an actor residing in municipality $m$ to travel to municipality $m'$ on a workday is then:
\begin{equation}\label{eq:gravweightsfreq}
    w_{m,m'}^{\mathrm{freq}} = \frac{P(m) P(m')}{R},
\end{equation}
where $P(m)$ is the population of municipality $m$, and $R$ is the distance between municipalities.
For incidental trips to visit family or for recreation, weights were used with $G = \frac{1}{7}$, and the square root of distance was used:
\begin{equation}\label{eq:gravweightsinc}
    w_{m,m'}^{\mathrm{inc}} = \frac{P(m) P(m')}{7 \sqrt{R}}
\end{equation}

\noindent Schedules for students and working adults were generated using the frequent trip weights on workdays, as they regularly commute to work or school in the same municipality. During weekends, these groups shift to incidental trips. In contrast, individuals from groups without work or school commitments always had their trips sampled with the incidental trip weights. 

\vspace{2mm}
\noindent The resulting average time spent outside the home municipality per demographic is shown in Tab. \ref{tab:demographic_groups}, with values similar to our assumptions. Furthermore, we know that 48$\%$ of students do not live in their home municipality \cite{Kences2020}, and $62\%$ of working adults do not live in the municipality where they work \cite{CBS2017}. These datasets were used to estimate how many students and working adults need to travel for work or school. The validity of actors' movements was checked against national surveys done by the Statistics Netherlands (CBS) \cite{CBS2017} and the Netherlands Institute for Social Research (SCP) \cite{SCP2022}. An average of $16.1\%$ of time is spent outside the home municipality in the model, which is lower than the observed $16.6\%$, however, uncertainties in the observed values are large \cite{SCP2022}. 

\vspace{2mm}
\noindent The use of gravity model weights, rather than more sophisticated data-driven methods, is justified by an outlier study \cite{simini2021}. Hospital admission data during the first COVID-19 wave in the Netherlands were reproduced by the model using both gravity model weights, and weights derived from human mobility data. No large difference in accuracy was found between the two approaches. From this comparison, it appears that much of the heterogeneity in real-life data was preserved through other means, including the incorporation of highly heterogeneous demographic compositions, data-informed travel schedules, and the mixing described in the following section.

\vspace{2mm}
\noindent Finally, to improve robustness of the results, five separate mobility networks were generated. Simulations in this study were conducted on every network before aggregating the results. The 75 simulations that were done per municipality per demographic group in generation of Figs. \ref{fig:risk_map} and \ref{fig:transmission_map} can therefore be interpreted as 15 simulations per municipality, per mobility network.

\subsection*{Step 4: Social mixing of the actors}

Heterogeneous mixing patterns of human networks were reproduced using several mixing matrices derived from real contact data in different contexts \cite{premetal2017}. These mixing matrices were used to sample contacts for each actor at an hourly basis. The sampled contacts matched the actor's demographic group and location, following the method outlined in the POLYMOD study \cite{mossongetal2008}. For example, an actor belonging to the working adults demographic group who was away from home at 3 PM would have used the `work' mixing matrix. Sampling from this matrix ensured that the actor was unlikely to encounter primary school children at work. In total, four $(11 \times 11)$ mixing matrices were used: `home', `work', `school', and `other', where the matrix element $C_{ij}$ represents the number of contacts an actor in group $i$ has with an actor in group $j$. The specific mixing matrices applied are shown in Tab. \ref{tab:demographic_groups}, categorised into Daytime mixing (between 08:00 and 18:00) and Nighttime mixing (all other times).  

\subsection*{Step 5: Pathogen transmission mechanism}
 
Pathogen transmission was simulated according to a SEIR model. In such a model, a susceptible actor (S) could become exposed, i.e. infected but not yet infectious (E) to the pathogen through contact with an infectious actor (I) until they became recovered and immune (R). The infection rate or (also called force of infection) $\lambda$ on an actor was calculated as the number of contacts with infectious actors multiplied by $\beta$, the probability of infection per contact. Thus, infectious actors exerted an infection pressure on each susceptible actor weighted by their mixing. The force of infection $\lambda$ on a susceptible actor from demographic group $g$, at time $t$, in municipality $m$, was calculated using the following equation:  
\begin{equation}\label{eq:infpressure}
    \lambda(g,m,t) = \beta \cdot s(t) \cdot \sum_{g'} n_{g,g'} \cdot \frac{I(g',m,t)}{N(g',m,t)},
\end{equation}
where $g'$ and $s(t)$ represent the demographic groups and the actor's daily activity cycle, $n_{g,g'}$ is the expected number of contacts for an actor in group $g$ with actors from group $g'$, and $\frac{I(g',m,t)}{N(g',m,t)}$ represents the fraction of infectious actors in demographic group $g'$ present in municipality $m$ at time $t$ \cite{dekker2023}. The actor was considered exposed if a random number chosen from a uniform distribution with a unit interval was less than $\lambda(g,m,t)$.

\vspace{2mm}
\noindent To simulate the spread of a novel respiratory pathogen, disease-specific parameters had to be selected. In this study, all simulations were performed using parameters loosely based on influenza A and SARS-CoV-2. The specific parameter values are given in Tab.~\ref{tab:pathogen_params}. Susceptibility to the pathogen was assumed to be equal across demographic groups. The latent and infectious periods represent the transition times to go from E to I, and I from to R, respectively. Since these periods vary among individuals, they were drawn from Weibull distributions with means of two and five days. The parameters of the Weibull distributions were chosen such that approximately half of the infectivity occurred three days after exposure. The incubation period, defined as the time from exposure to symptom onset, was set to three days. Consequently, if self-isolation began at symptom onset, approximately half of all infections could have been prevented. For adherence rates less than unity in both the self-isolation and mobility restriction measures, the self-isolating actors were chosen randomly. 
\begin{table}[!h]
\centering
\renewcommand{\arraystretch}{1.4} 
\setlength{\tabcolsep}{4pt}

\begin{tabular}{|p{7.5cm}||Q{7.5cm}|}
\hline
\rowcolor{white}
\textbf{Parameter} & \textbf{Details} \\ 
\hline
\hline
Probability of infection ($\beta$) & 0.135 per contact \\
\hline
Latent period ($T_E$) & 2 days ($90\%$ range: 1.2-3.5 days)\\
\hline
Infectious period ($T_I$)& 5 days ($90\%$ range: 1.6-8.9 days)\\
\hline
Incubation period ($T_S$)& 3 days \\
\hline
Percentage of infectious period after symptom onset & $47\%$ ($90\%$ range: $31\%$, $64\%$)\\
\hline
\end{tabular}
\caption{Transmission parameters for the respiratory pathogen.}
\label{tab:pathogen_params}
\end{table}

\vspace{2mm}
\noindent Finally, $\mathcal{R}_0$ is defined as the expected number of secondary infections generated by a typical infectious individual introduced into a wholly susceptible population. In general, $\mathcal{R}_0$ is the spectral radius of the next-generation operator obtained by linearising a dynamical system around the disease-free equilibrium \cite{diekmann2013}. For our actor-based model, which features heterogeneous contact structures across households, schools, workplaces and municipalities, such a linearised representation is not available. Instead, we follow the renewal-equation approach developed in demographic theory (Euler–Lotka) and adopted for infectious disease dynamics \cite{diekmann2013, yan2008}. 

\vspace{2mm}
\noindent The value of $\mathcal{R}_0$ for our model is $\approx3.6$. The calculation of $\mathcal{R}_0$ can be found in SI F. We stress that its value is not imposed {\it a priori\/}, but is implied by the combination of the simulated early growth rate and the assumed distributions of the latent and infectious periods.
\begin{figure*}[!h]
\centering
\includegraphics[width=\textwidth]{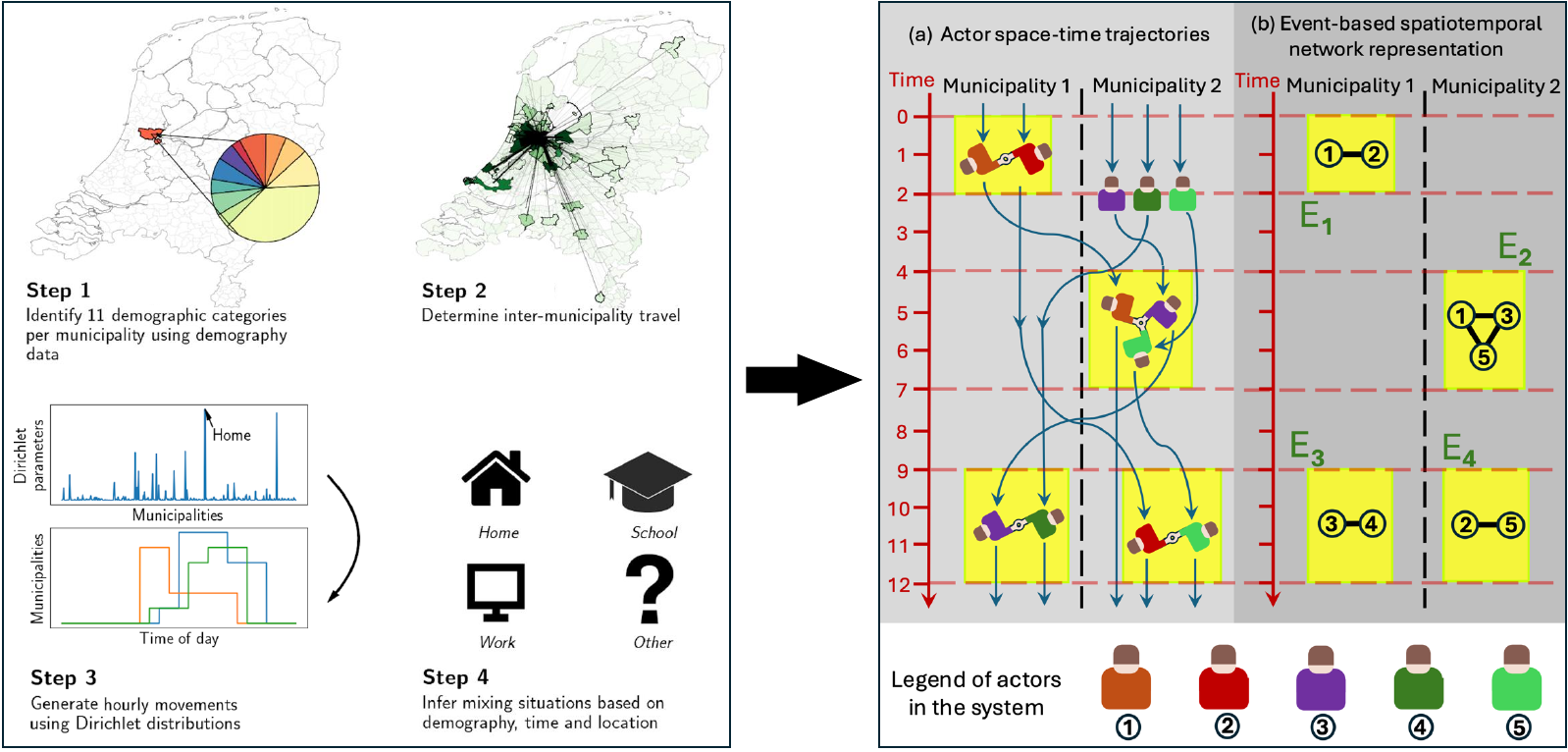}
\caption{{\bf Model flowchart describing the construction of the dynamic contact network amongst actors}. Step 1: actors are distributed across municipalities reflecting the demographic compositions at both the national and the municipality level. Steps 2 and 3: actors move from municipality to municipality based on a travel schedule informed by real-life data at hourly resolution. Step 2 shows, in darkness scale, how many actors from which municipality are present in a given municipality (in this case municipality of Amsterdam) at a given hour, while step 3 shows stochastically generated movement patterns for individual actors. Step 4: Once the actor composition is known in a given municipality at any hour, then they interact with amongst each other in four situations (home, school, work or other). After all this is implemented, a full actor spatiotemporal trajectory is generated [panel (a), right box] over a period of time. The spatiotemporal trajectory is encoded in the computer in terms of events that describe actor interactions [panel (b), right box]. Events, tagged E$_1$-E$_4$  represent the specific spatiotemporal anchor points for actor interactions. The event-based dynamic spatiotemporal representation yields a significant speed-up in comparison to a corresponding agent-based simulation. Time (starting arbitrarily at zero) is measured in units of an hour, with municipalities as the lowest units of space. The left box has been adapted from Ref. \cite{dekker2023}. \label{fig:model_schematics}}
\end{figure*}

\section*{Results}

\subsection*{SEIR model on dynamic network at population scale}

An actor-based model was used to simulate an epidemic in the Netherlands caused by a novel respiratory pathogen, whose characteristics are based on respiratory viruses such as influenza A and SARS-CoV-2. In the model, individual municipalities were chosen as the smallest unit for spatial resolution, and time resolution was chosen to be an hour. Actors at 1:100 population ratio were then distributed across municipalities such that they reflected the demographic compositions at both the national and the municipality levels (Methods, step 1). These actors moved from municipality to municipality based on a travel schedule informed by real-life data at hourly resolution, and their hourly locations and interactions (events \cite{dekker2022,dekker2022b}) were tracked in time (Methods, steps 2 and 3). Once the actor composition of every municipality per hour was obtained, they mixed with each other in four different situations (or settings): home, school, work and other (Methods, step 4). This generated a nationwide {\it stochastic\/} and {\it dynamic\/} (spatiotemporal) contact network at hourly resolution (see Fig. \ref{fig:model_schematics}). On this dynamic network we simulated pathogen transmission using a Susceptible-Exposed-Infectious-Recovered (SEIR) model (Methods, step 5).

\subsection*{Risk mapping an early epidemic}

We considered a scenario where, on day zero, a respiratory pathogen is introduced into five actors of a specific demographic group residing in a given municipality (hereafter, the ``seed municipality''). From then on, facilitated by cross-municipality mobility and local mixing, the epidemic takes off. Table~\ref{tab:pathogen_params} summarises the pathogen's disease characteristics, which represent typical traits of directly transmitted pathogens well-adapted for human-to-human transmission in a susceptible population \cite{fraser2004}.
\begin{figure}[!h]
\centering
\includegraphics[width=\linewidth]{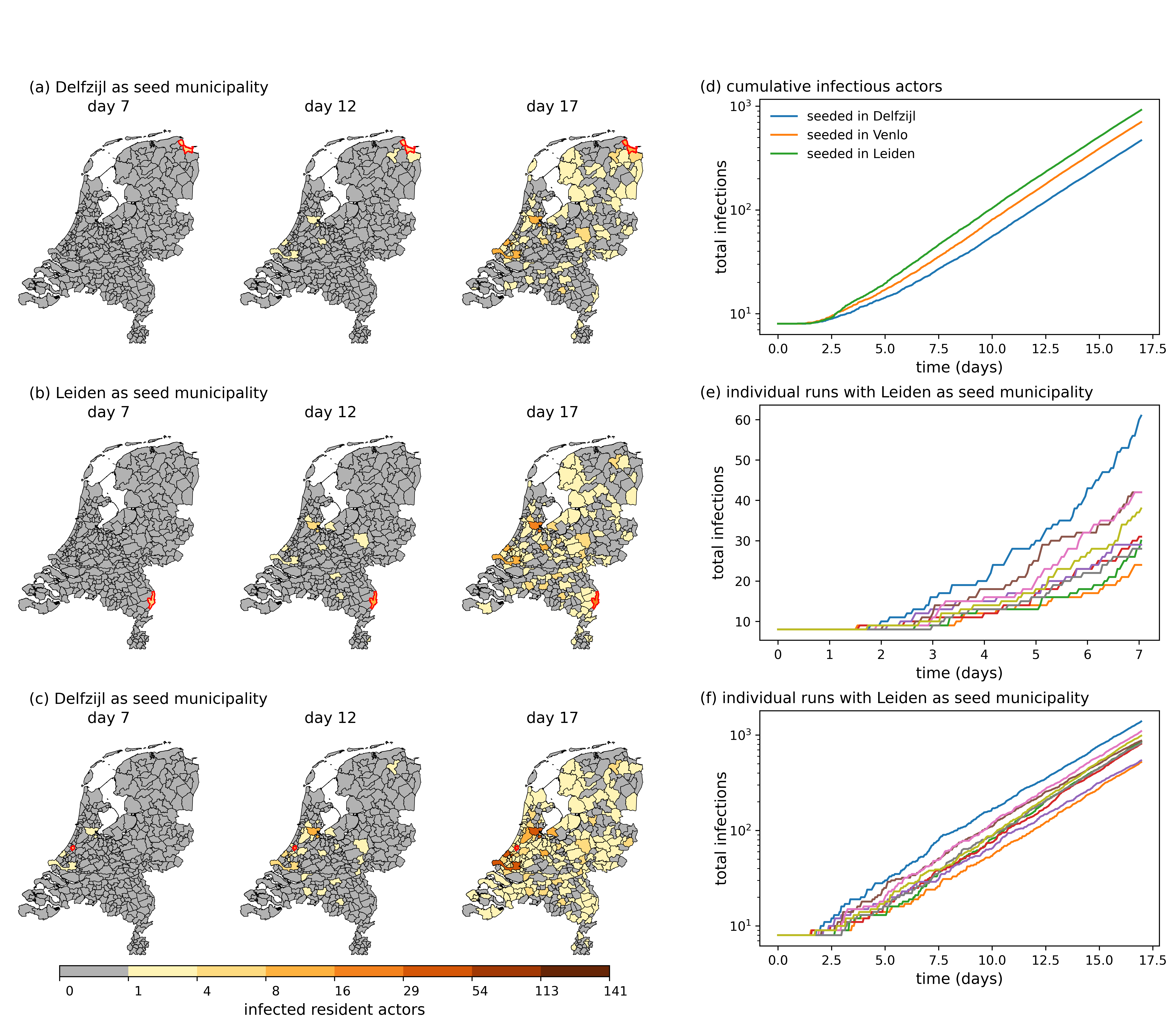}
\caption{The ``infection intensity maps'' of the Netherlands, i.e., cumulative number of infections over 17 days, following pathogen introduction in five working adults of Delfzijl (a),  Venlo (b), and Leiden (c) on day zero. For all three seed municipalities we also show the total cumulative infectious actors in panel (d) over the full 17 days. The cumulative number of infections and geographic spread are significantly greater for Leiden as the seed municipality than for Delfzijl and Venlo, due to its proximity to Amsterdam, the Hague, and Rotterdam, three large cities in the Netherlands. Results are averaged over 75 simulation runs. Stochastic variability across simulation runs are showcased for Leiden as the seed municipality, for early (e) and somewhat later (f) times (see text for details).}
\label{fig:two_municipalities}
\end{figure}

\subsubsection*{Risk mapping by municipality of introduction}

Figure~\ref{fig:two_municipalities} showcases the ``infection intensity map'': the state of the epidemic in terms of the cumulative number of infectious (I) actors per municipality over the first 17 days of the outbreak, caused by the introduction of the pathogen in five working adults in a given municipality on day zero. Three scenarios are exemplified, with seed municipalities: Delfzijl, located in the sparsely populated Northeast [panel (a)], Venlo, a small city in the East, relatively far away from major urban centres [panel (b)], and Leiden, situated near the major cities Amsterdam and Utrecht [panel (c)]. Within about two weeks, pronounced differences emerge in the distribution of infectious actors across municipalities: infections are considerably fewer, and the geographic spread is slower when the seed municipality is Delfzijl, compared to when the seed municipality is Leiden. The cumulative number of infections in the whole country in the first couple of days for Venlo and Delfzijl as seed municipalities also lags behind that for Leiden. However, within about one week, the exponential growth rates of these cities catch up with that for Leiden as the seed municipality [panel (d)]. Finally, considerable variations exist among simulation runs seeded in a given municipality [shown for Leiden in panels (e-f)]. This holds for all municipalities, and is caused by stochastic effects playing a major role in the early stages of the epidemics. Of note, we ensured the robustness of the infection intensity maps in Fig. \ref{fig:two_municipalities} by testing for their sensitivity in SI A to the number of initially seeded actors (using eight working adults instead of the five used in Fig. \ref{fig:two_municipalities}).

\begin{figure}[hpt!]
\centering\includegraphics[width=\linewidth]{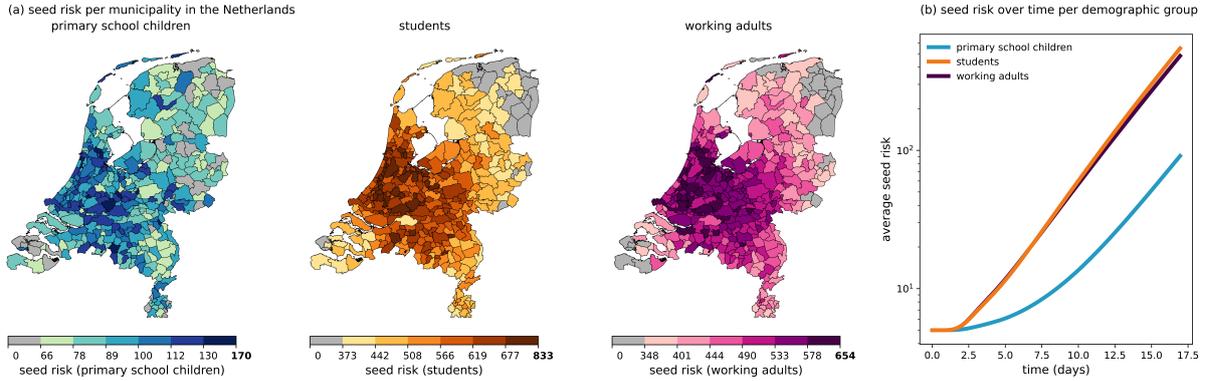}
\caption{Seed risk characterised at municipality-resolution for a novel respiratory pathogen in the Netherlands. (a) Shown in colour coding is the seed risk posed by a municipality, defined as the cumulative number of national infectious actors after 17 days following the introduction of the pathogen into five infectious actors in that municipality on day zero (see main text). The average across municipalities is displayed in panel (b).
In both panels results are shown for pathogen introduction by primary school children (left), students (middle), and working adults (right).
Results are averaged over 75 simulation runs per seed municipality and demographic group of introduction.}
\label{fig:risk_map}
\end{figure}

\vspace{2mm}
\noindent Using the information contained in Fig.~\ref{fig:two_municipalities} we constructed quantitative seed risk maps for all 355 municipalities in Fig.~\ref{fig:risk_map}. These maps represents the cumulative total number of infectious actors (i.e. summed over all municipalities) after 17 days following introduction of the pathogen in each municipality on day zero. The seed risk maps reveal that there are considerable spatial variations in risk, as well as a strong dependence of the risk on the demographic group into which the pathogen is introduced. Furthermore, seed risk is typically much higher when the seed municipality is near large well-connected cities in the west of the Netherlands. These large cities act as the ``core group'' for pathogen transmission. This phenomenon has been observed and modelled in the context of compartmental and metapopulation models \cite{lau2020, madden2024}, and our results show that these phenomena can also be found in dynamic network models with data-informed contact patterns and mobility. Below we elaborate on this observation by tracking where pathogen transmissions occur. Of note, we tested the sensitivities of the seed risk maps and the seed risk scores to the time horizon length (14-, 17- and 21-days) in SI B. These tests show that the 17-days seed risk scores are robust.

\begin{figure*}[!h]
\centering
\includegraphics[width=\linewidth]{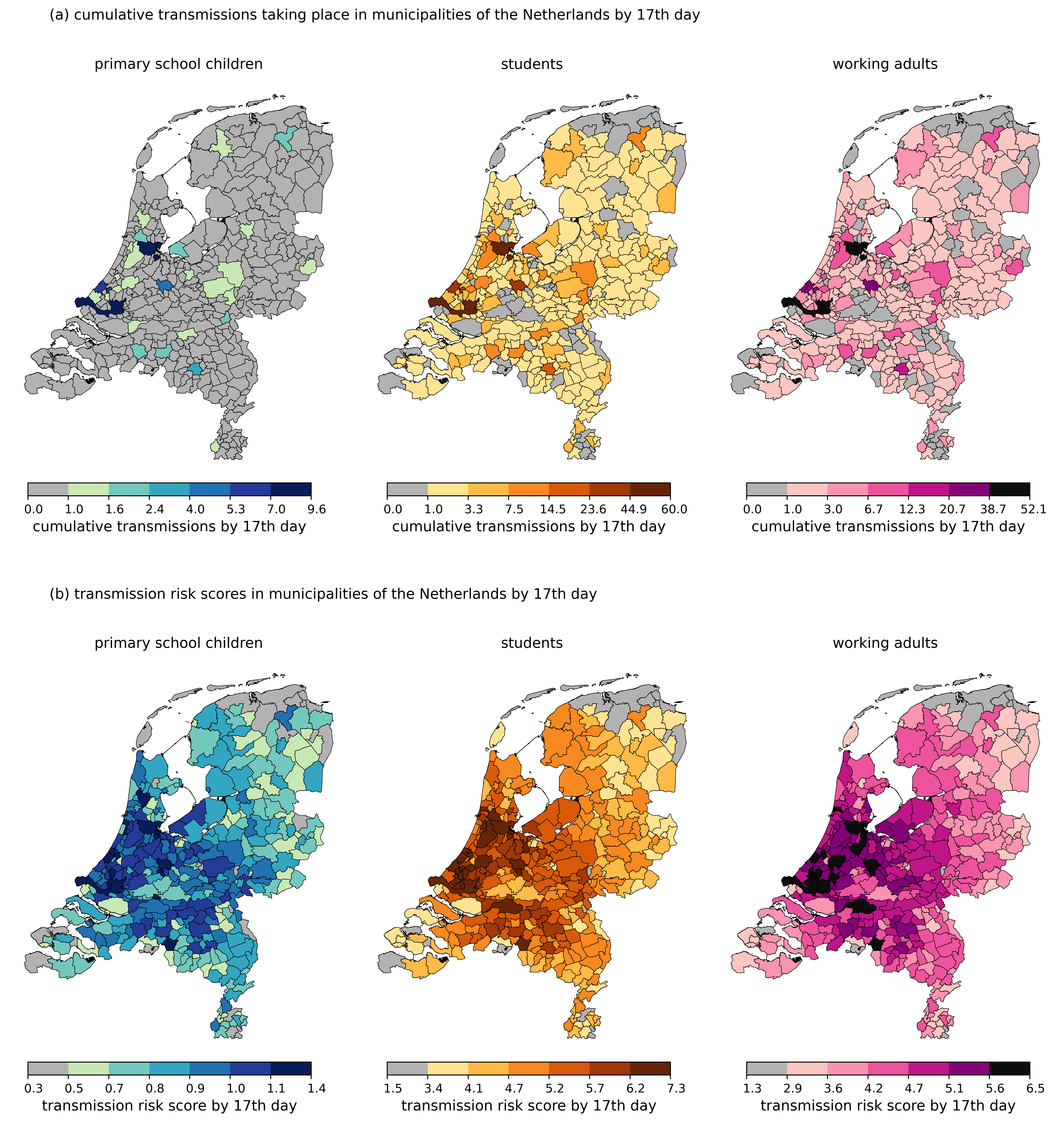}
\caption{Municipality-resolved transmission maps, and transmission risk score maps, stratified by the demographic group into which the pathogen is introduced. (a) Transmission risk map defined by the cumulative number of transmissions taking place within each municipality on day 17, upon randomly introducing the pathogen into five actors of a demographic group in a municipality on day zero, calculated as follows. The number of cumulative transmissions taking place in a municipality is weighted by the probability $P$ of the epidemic originating in one municipality, calculated as the fraction of actors from the seeding demographic group living in that municipality; i.e., $P(\mathrm{municipality\ } m\ \mathrm{ being\ the\ seed}) = \frac{n(g,m)}{N(g)}$, where $n(g,m)$ is the number of actors of group $g$ in municipality $m$, and $N(g)$ is the national number of actors in group $g$ (see further the Transmission section in Methods). (b) Correspondingly, the risk score map (transmissions taking place per capita in a municipality). See definition of risk score in the main text. Results are averaged over 75 runs per municipality and demographic group.}
\label{fig:transmission_map}
\end{figure*}

\subsubsection*{Risk mapping by transmission intensity}

Transmission of the pathogen between a susceptible and an infectious actor, who are potentially residents of two different municipalities, can take place in a third municipality where both actors have travelled to. This observation naturally raises a question relevant for epidemiological dynamics: could there exist a group of municipalities that act as primary multipliers for pathogen transmission, which is often referred to as the core group driving the epidemic? To answer this question, we simulated epidemics upon introduction of the pathogen into a demographic group residing in one municipality at a time, and tallied the resulting transmission events cumulatively over 17 days by the municipality where the transmission took place, weighted by the municipality-based population sizes of the demographic groups into which the pathogen was introduced. This results in a municipality-resolved ``transmission map'' stratified by the demographic group into which the pathogen is introduced [Fig.~\ref{fig:transmission_map}, panel (a)]. Consistent with the risk map in Fig.~\ref{fig:risk_map}, of the three demographic groups shown in Fig.~\ref{fig:transmission_map}, the total number of infections after 17 days is highest when the pathogen is introduced in students and lowest when introduced into primary school children.

\vspace{2mm}
\noindent Importantly, transmission events are distributed highly unevenly over municipalities, with many more taking place in heavily populated municipalities than in smaller ones (national mean for municipality population is approximately $43,000$). Especially the four largest municipalities, Utrecht, Amsterdam, Rotterdam and The Hague in the central west part of the country have a disproportionate number of attributed transmissions when compared to their population size, as shown in Tab.~\ref{tab:bigcitypopulations}. This is true both in terms of absolute numbers [Fig. \ref{fig:transmission_map}, panel (a)] and transmission risk scores, defined as cumulative number of transmissions taking place in a municipality, per 1,000 residents in that municipality [Fig. \ref{fig:transmission_map}, panel (b)]. These results are consistent with those of Figs.~\ref{fig:two_municipalities}-\ref{fig:risk_map}, and we conclude that densely populated part in the west part of the Netherlands (especially the large cities), act as core groups of epidemic spread in the early stages of an epidemic.
\begin{table}[hpt!]
\centering 
\begin{tabular}
{|p{0.11\linewidth}||
                W{0.095\linewidth}|
                Q{0.15\linewidth}|
                W{0.155\linewidth}|
                Q{0.12\linewidth}|
                Q{0.17\linewidth}|}
\hline
\rowcolor{white}
Municipality & Population size & Percentage of population & Percentage of transmissions & Transmission Risk score & (90\% range)\\
\hline
\hline
Rotterdam & 641,000 & 3.9\% & 4.7\% & 7.00 & (6.27 - 7.83)\\
\hline
Amsterdam & 860,000 & 5.2\% & 6.2 \% & 6.97 & (6.11 - 8.04)\\
\hline
Utrecht & 350,000 & 2.1\% & 2.5\% & 6.74 & (5.75 - 7.91)\\
\hline
The Hague & 536,000 & 3.2\% & 3.7\% & 6.59 & (5.70 - 7.57)\\
\hline
\end{tabular}
\caption{Population data (2019) and transmission risk scores in four highest-populated municipalities in the Netherlands, ranked by their risk scores, when the pathogen is introduced in the student population. See definition of risk score in the main text. The full list can be found in SI C.}
\label{tab:bigcitypopulations}
\end{table}

\vspace{2mm}
\noindent We tested the sensitivities of the transmission risk maps and the transmission risk scores to the time horizon length (14-, 17- and 21-days) in SI C. These tests demonstrate that the 17-days seed risk scores are robust. The transmission risk scores for the four most populous municipalities are noted in Tab. \ref{tab:bigcitypopulations}.

\subsection*{Impact of behavioural changes and targeted interventions}

One of the biggest challenges in epidemiological modelling is that the dynamics of an epidemic cannot be untangled from human behaviour \cite{funk2015, walker2020, bedson2021, brindal2022, ryan2024}. Behavioural changes can be a response to a targeted intervention measure, but can also arise spontaneously. Moreover, human behaviour is adaptive; depending on the available information regarding the state of the epidemic, e.g., the infection intensity, residents can adapt their travel and mixing behaviour to avoid infection risks. The degree of adaptivity can vary a lot: not only across different demographic groups, but, as the SARS-CoV-2 pandemic has clearly demonstrated, also across socio-economic status \cite{jay2020, chang2021, lee2022}. To make matters even more complicated, adaptive behaviour, being dependent on the state of the epidemic, is time-dependent. Our hybrid actor-based model playing out on a dynamic network is eminently suited to realistically capture the impacts of granular-level behavioural changes on epidemic dynamics on the one hand, and simultaneously scaling it up to the population level on the other.

\vspace{2mm}
\noindent We explored the impacts on epidemic dynamics due to (1) symptom-based self-isolation as an example of adaptive behaviour, and (2) reduced mobility across the borders of the core group of municipalities as an example intervention measure. For both cases we simulated the epidemic dynamics upon introducing the pathogen into a five working adult actors residing in one municipality at a time, and under various scenarios of behavioural change, tallied the resulting transmission events cumulatively over 17 days by the municipality where the transmission took place, weighted by the municipality-based population sizes of the demographic groups into which the pathogen is introduced. In both the self-isolation and mobility interventions we assumed that the behavioural changes are in effect from day zero onwards. Hence, this corresponds to a best-case scenario with no detection, policy or behavioural delays.

\subsubsection*{Symptoms-based self-isolation}\label{sec:selfisolation}

\noindent Throughout we used a latent period of approximately 2 days (90\% range: 1.22-3.50 days), an infectious period of approximately 5 days (90\% range: 1.63-8.92 days), and an incubation period of 3 days (Tab.~ \ref{tab:pathogen_params}). From these parameters, it follows that 47\% of the infectious period occurs after the onset of symptoms, implying that if we assume self-isolation begins immediately upon symptom onset, we expect that approximately half of the transmissions can potentially be prevented. In order to incorporate variability and realism in our exploration, we considered scenarios with respect to the fraction of individuals that adhere to self-isolation. 

\begin{figure}[hpt!]
\centering
\includegraphics[width=0.95\linewidth]{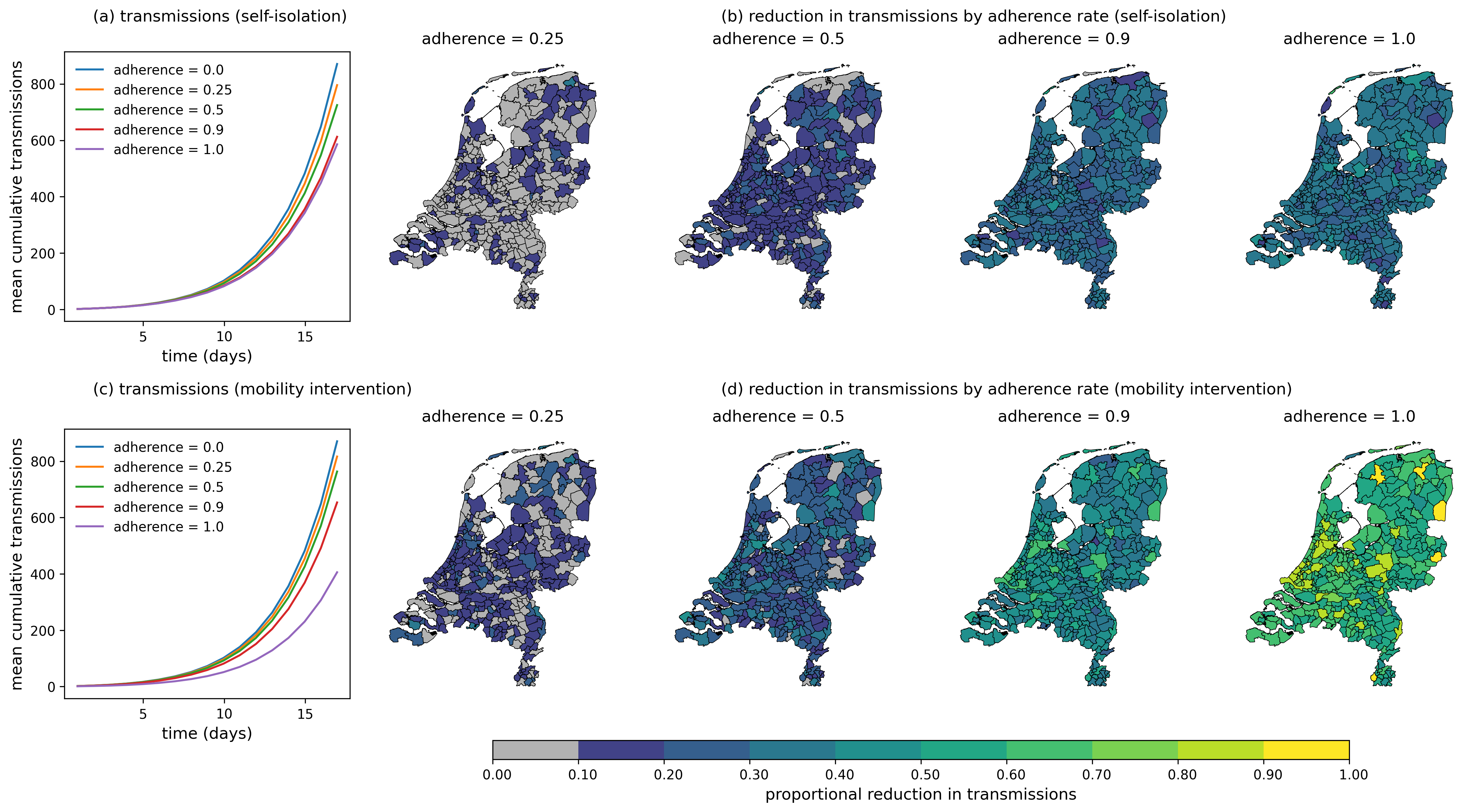}
\caption{Impact of interventions on pathogen transmission. (a) and (b): symptoms-based self-isolation; (c) and (d): movement restrictions to and from municipalities with more than $100,000$ inhabitants). See main text for details. (a) and (c): mean cumulative national number of transmissions; (b) and (d): reduction of transmissions relative to no intervention for four different adherence rates. Actors are randomly selected to be adherent. The seed actors are working adults. Results are averaged over 25 simulations per municipality per adherence rate. The 90\% range of the transmissions for various adherence rates can be found in SI D (they have been left out of the plots in order to avoid cluttering them).}
\label{fig:selfisolation}
\end{figure}

\vspace{2mm}
\noindent In Fig.~\ref{fig:selfisolation} we show the transmission risks for varying fractions of adherence. Adherence is defined as the fraction of symptomatic actors that choose to self-isolate. Panel (a) shows mean national cumulative transmissions for five different adherence fractions (the reduction in cumulative transmissions after 17 days for each adherence fraction is presented in Tab. \ref{tab:interventionreductions}). Although the cumulative number of transmissions is reduced for each adherence rate, the measured effect for adherence rates of 0.25 and 0.5 is relatively modest. The corresponding spatial distribution of  {\it proportional\/} reductions of transmission risk is shown in Fig.~\ref{fig:selfisolation}b. While reductions do vary across municipalities, we find no clear relation between the location and the amount by which transmission is reduced.

\subsubsection*{Mobility reduction across the borders of the core group of municipalities}

Based on the risk maps of Fig. \ref{fig:transmission_map}, we explored scenarios built around selectively restricting mobilities to and from the core group of municipalities; specifically, across borders of municipalities with resident populations exceeding 100,000. As an illustrative experiment, we choose to implement the mobility restriction from day zero. Apart from changing mobility patterns, simulations are set up as earlier. In other words, actors moved and mixed in the same way as before, except that they cannot cross the borders of municipalities with populations exceeding 100,000.

\vspace{2mm}
\noindent Figure \ref{fig:selfisolation} shows the impact of adherence on the cumulative number of transmissions. Adherence is defined as the fraction of individuals who respect the borders of some municipalities being closed. Non-adhering individuals will travel exactly as they did before (both into and out of the closed municipalities), while adhering individuals will stay in their home municipality. Fig.~\ref{fig:selfisolation}a shows that transmissions are reduced at each level of adherence (see also Tab.~\ref{tab:interventionreductions}).
The {\it proportional\/} transmission reduction maps in Fig.~\ref{fig:selfisolation}b show that reduction is skewed towards densely populated municipalities. Notably, the map for adherence rate 0.9 shows strong reductions in densely populated municipalities and lower reductions in less populated municipalities in the east and south. Interestingly, reduction of mobility across the borders of core group appears most effective for adherence rates approaching unity, with transmission reduction doubling between adherence rates of 0.9 and unity. We attribute this strong dependence on adherence to the fact that without any measures most transmissions happen in heavily populated municipalities (see Fig.~\ref{fig:transmission_map}) and that at an adherence rate of unity, the populous municipalities are effectively isolated.
\begin{table}[hpt!]
\centering  
\begin{tabular}{|p{0.16\linewidth}||W{0.27\linewidth}|Q{0.31\linewidth}|}
\hline
\rowcolor{white}
\multirow{2}{*}{Adherence level} 
  & \multicolumn{2}{c|}{Reduction of cumulative transmissions} \\
\cline{2-3}
  & Self-isolation & Mobility restriction \\
\hline
\hline
0.25 & 7.8\%  & 13.5\%   \\
\hline
0.5  & 16.7\% & 27.4\%   \\
\hline
0.9  & 30.0\% & 49.0\%   \\
\hline
1.0  & 33.6\% & 71.0\%   \\
\hline
\end{tabular}
\caption{Proportional reduction in cumulative transmissions after 17 days by self-isolation and mobility restrictions, by adherence rate. See text for details.}
\label{tab:interventionreductions}
\end{table}

\vspace{2mm}
\noindent By comparing both interventions to one another using Fig. \ref{fig:selfisolation} and Tab. \ref{tab:interventionreductions}, we find that the mobility restriction measure performs better at every adherence rate, nearly doubling reductions compared to the self-isolation measure. Notably, the mobility restriction measure improves from 49\% to 71\% between adherence rates of 0.9 and unity, whereas between the same rates, the self-isolation measure barely improves. From these results, it appears that completely isolating the core group of municipalities can be very effective in reducing the growth rate of an early stage epidemic, for the natural history parameters of our respiratory pathogen (incubation and latent periods, infectious period, transmissibility).

\vspace{2mm}
\noindent Finally, as a comparison to Fig. \ref{fig:selfisolation} the impact of movement restrictions to and from municipalities with more than 150,000 inhabitants is shown in SI E.

\section*{Discussion} 
 
\noindent Using the Netherlands as a case study, we performed a high-resolution analysis of the early outbreak dynamics of a respiratory pathogen, enabling identification of key transmission hubs and quantification of municipality-specific risks. Specifically, for a given set of pathogen parameters and mobility patterns, we have developed quantitative risk scores for all municipalities (Tabs. SI.1-SI.3). The risk score provides a measure of the expected cumulative number of cases, relative to population size, thereby integrating pathogen characteristics with human demography and mobility. This can be used in early outbreaks to provide a near real-time tool for local public health risk assessments. 

\vspace{2mm}
\noindent At the coarse end of the aggregation scale, e.g., traditional metapopulation models, such as those recently applied to influenza~A and SARS-CoV-2 \cite{colizza2007, balcan2009, belik2011} and earlier to measles \cite{bolker1996, grenfell1998, grenfell2001, bjornstad2002, lau2020, madden2024}, have demonstrated the role of mobility-driven spread and persistence. Such models can capture the general characteristics of broad spatial transmission patterns. However, they lack the ability to represent the realistic, fine-scale, stochastic nature of human interactions that dominate the early outbreak dynamics as they often assume deterministic transmission, homogeneous mixing within subpopulations, and highly simplified mobility rates. In contrast, models at the fine end of the scale are often difficult to parametrise realistically, and remain difficult to generalise and scale up. Our hybrid approach combines the spatial structure and demographic stratification of a metapopulation framework with a dynamically generated, data-informed human contact network. By integrating empirical mobility patterns into stochastic contact formation, we achieve both generalisability and realism. We show how municipality-level risk scores emerge from the interaction between mobility-driven network dynamics and epidemiological processes. Moreover, our risk mapping can be applied in the early stages of an epidemic to determine the expected spatio-temporal range of disease spread, and to anticipate how an infection will propagate following a focussed introduction. Such analyses extend beyond static risk assessments, allowing for real-time evaluation of transmission hotspots and potential control measures. 

\vspace{2mm}
\noindent Our study holds a number of implications for public health policy and control for directly transmitted respiratory infections. In fact, our analyses underscore that especially in the early stages of an unfolding epidemic or pandemic, the incidence and risk of infection can be very heterogeneous in space, as can the speed with which the pathogen will spread. For the parameters chosen here, reflecting an acute respiratory pathogen, containment after a focussed introduction will be very difficult after two to three weeks in the densely populated western part of the Netherlands, but may still be possible in more remote, less well-connected municipalities. Clearly, the relative time scale on which containment is still possible will vary between pathogens, with more transmissible pathogens that have shorter generation times and relatively long asymptomatic periods leaving less time for containment than less transmissible pathogens with longer generation times and short asymptomatic periods \cite{fraser2004}.

\vspace{2mm}
\noindent We discuss several simplifying assumptions and future extensions. First, susceptibility and infectiousness in the model are uniform across demographic groups. While this assumption allows us to focus on spatial and temporal transmission heterogeneity, real-world pathogens exhibit age- and comorbidity-dependent susceptibility (e.g., \cite{davies2020, haug2023}). The model is age-stratified, allowing future analysis to incorporate age-specific attack rates and immunological factors that affect susceptibility. Second, the model currently does not include explicit household structures or workplace-specific contact networks. While the use of setting-specific mixing matrices partially accounts for differential contact patterns, more detailed modelling of structured interactions within households, schools, and workplaces --- if such information is available --- will improve realism. This will open up avenues to study the impact of local and tailored interventions in households and schools, including various realistic contact tracing strategies. For the Netherlands, we envisage model extensions that incorporate the school-community network (e.g., \cite{munday2024}) and thereby investigate the impact of local interventions such as school closures in more detail \cite{rozhnova2021}. This is an important avenue for future development, as it will enable the study of the extent to which details of the contact network affect the population-level transmission dynamics. Third, the current use of municipality-level resolution for mobility is too coarse-grained to fully capture within-municipality travel patterns. While our approach represents an improvement over static networks, finer spatial resolution (e.g., neighbourhood-level travel data) could further refine the identification of high-risk transmission locations. Similarly, gravity model weights appear to be less heterogeneous in space than weights sourced from realistic data \cite{dekker2023}. Therefore, the resulting mobility in this study should be seen as a lower bound for real-life heterogeneity. Using mobility data in the future could provide us with a much more realistic view of epidemic dynamics, and their resulting transmission and seed risks. Furthermore, while our mobility model is calibrated to available data, it does not account for behavioural adaptations in response to disease spread. Empirical studies of SARS-CoV-2 suggest that human mobility patterns change dynamically in response to perceived risk (e.g., \cite{chan2020, yabe2020, kraemer2020, holtz2020}), and these can potentially be integrated in a data-driven manner in future iterations of the model (see also \cite{dekker2023}). Fourth, reactive behavioural adaptations (such as self-regulation, inclusive self-isolation) can be incorporated in a more realistic manner than hitherto possible, e.g., by assigning specific (probabilistic) behavioural codes to individual actors. 

\vspace{2mm}
\noindent Of course, the finer the resolution (population, space, time, mobility, and local mixing) in the model, the more the computational burden increases. We do not know where the reasonable-to-unreasonable boundary lies for computational burden. So far, without having optimised for efficiency, we have been able to experiment with a 1:25 population ratio, which means about 680,000 actors, with spatial and temporal resolutions as in the current study. 

\vspace{2mm}
\noindent Concluding, we found that the early epidemic dynamics of novel respiratory pathogens is heterogeneous in the epidemiologically aspects that we have studied, already in the small, densely populated, and well-connected population of the Netherlands. This suggests that such heterogeneities may be even larger and more impactful in larger countries and regions that are more sparsely connected by human mobility, opening up opportunities for optimal locally-tailored public health response for containing the pathogen. This is our main direction for future research.

\vspace{2mm}
\noindent {\bf Author contributions.} MR, MvB and DP contributed to developing the key concepts presented here. MR performed simulations. MR, MvB and DP contributed to writing the manuscript.

\vspace{2mm}
\noindent {\bf Competing Interests.} The authors have no competing interest to declare.

\vspace{2mm}
\noindent {\bf Data availability.} The demography, residency and municipality data are publicly available through the Statline tool of the Central Bureau of Statistics of the Netherlands (CBS) \cite{CBSstatline}. Data used for mixing have been taken from Refs. \cite{mossongetal2008} and \cite{premetal2017} (see Methods, Step 4). Data for reproduction of the figures can be found in an archived Github repository \cite{githubrepo2025}. 

\vspace{2mm}
\noindent {\bf Code availability.}
The source code is publicly available in a GitHub repository at \url{https://github.com/matthijsromeijnders/covid-simulation-commsmedicine/}. Archived on Jan 5, 2026 on Zenodo \cite{githubrepo2025}. 

\vspace{2mm}
\noindent {\bf Acknowledgements.} This work has been funded in part by the project ``Real-time ruimtelijke data-gedreven modellering van uitbraken van infectieziekten'' with project number 10710062310008, financed by The Netherlands Organisation for Health Research and Development (ZonMw). We thank Sasha Teslya and other members of the infectious disease Modelling Group at the University Medical Center Utrecht for insightful feedback.

\end{document}